\documentclass[12pt]{article}
\usepackage{amsmath,amssymb,array}
\usepackage{graphicx}

\numberwithin{equation}{section}
\def\p{\partial}

\begin{document}

\begin{titlepage}
\renewcommand{\thefootnote}{\fnsymbol{footnote}}

\begin{center}

\begin{flushright}arXiv: 1002.4352\end{flushright}
\vspace{3.5cm}

\textbf{\Large{Improving the Excited Nucleon Spectrum \\[0.5cm] in Hard-Wall AdS/QCD}}\vspace{2cm}

\textbf{Peng Zhang} \\[0.5cm]

\textsf{E-mail: pzhang@bjut.edu.cn}\\[0.5cm]

\emph{Institute of Theoretical Physics, College of Applied Sciences, \\
      Beijing University of Technology, Beijing 100124, P.R.China}

\end{center}\vspace{1.5cm}

\centerline{\textbf{Abstract}}\vspace{0.5cm}
We show that the nucleon spectrum in a hard-wall AdS/QCD model 
can be improved by use of a relatively large IR cutoff. All of the spin-1/2
nucleon masses listed in PDG can be fit quite well within 11\%. The average
error is remarkably only 4.66\%.

\end{titlepage}
\setcounter{footnote}{0}

\section{Introduction}

The establishment of Quantum Chromodynamics (QCD) is a milestone of
theoretical physics. Because of the property of asymptotic freedom,
QCD is convenient to use in the process with large momentum transfer.
In the low energy region, however, QCD becomes strongly coupled,
and the perturbative calculation is not reliable. Here the effective
degrees of freedom are mesons and baryons. People have developed
various effective models to describe their properties.

Recently a new methodology, called ``AdS/QCD'' or more generally
``holographic QCD'', has been developed and intensively studied.
Actually it is originated from an old idea due to 't Hooft
\cite{tH}: The large $N$ limit of a gauge theory should be dual to
some string (or gravity) theory. This idea has been explicitly
realized by the AdS/CFT correspondence \cite{M, GKP, W}. The gauge
theory side is the $\mathcal{N}=4$ supersymmetric Yang-Mills, while
the other side is type IIB string theory in $AdS_5 \times S^5$.
Right after its discovery, people (mainly string theorists)
began to construct the gravity dual of QCD-like theories by reducing
the number of SUSY and breaking the conformal symmetry. Pioneer
works in this direction include \cite{KW}-\cite{V} and others. Based on
some basic setups in \cite{W9803, KMMW}, the authors of \cite{SS}
construct a semirealistic holographic dual of QCD. These are the
so-called top-down approach, which begin from some brane
configurations in string theory and try to use the field theory
living on these branes to mimic QCD. Motivated by these works,
\cite{EKSS, DP} develop a complementary bottom-up approach, which
has the advantage that it is related directly to some important
features of QCD and has much more flexibility to implement the
phenomenological studies. In this model the background metric is
chosen to be $AdS_5$, the fields involved are two gauge fields and a
bifundamental scalars, which are holographical duals of the chiral
symmetry currents and quark condensates respectively. The conformal
symmetry is broken by imposing an IR cutoff at the extra fifth
direction, so it is called as hard-wall model. The low-lying meson
spectrum can be reproduced quite well. To realize linear confinement
the authors of \cite{KKSS} also introduce a soft-wall model. Many
works have been done mainly in meson sectors, see e.g.
\cite{DaRold:2005vr}-\cite{Zuo}. There are also interesting
discussions about the relation between light-front dynamics and
AdS/QCD, see references in \cite{BdT}.

Baryons can also be studied in the hard-wall model, either by a
skyrmion-type soliton \cite{PW1, PW2}, or by introducing two Dirac
spinors in the bulk \cite{HIY}, see also \cite{MT, KKY}. In this
note, we will focus on the latter method. In this method, spin-1/2
nucleons are described as Kaluza-Klein modes of two 5d Dirac
spinors. The chiral symmetry breaking is introduced through a Yukawa
coupling between the 5d spinors and the scalar corresponding to the
quark condensate. By use of a IR cutoff $z_{\mathrm{IR}}\sim$ 5 GeV,
the mass of first three radial excitations of nucleon, i.e.
$N(939),\,N(1440),\,N(1535)$, can be reproduced well. However the
theoretical values for higher nucleon states are much larger than
the data. In this paper we show that by simply using a larger
$z_{\mathrm{IR}}$, all of the spin-1/2 nucleon masses listed in PDG
\cite{PDG} can be fit quite well within 11\%. The average error is
only 4.66\%.

This paper is organized as follows: In section 2 we will review some
basic facts about hard-wall model. In section 3 we will show our
numerical results about the nucleon spectrum. Section 4 is devoted
to the summary and some discussions.

\section{Mesons and Nucleons in hard-wall AdS/QCD}

In this section we will briefly review the hard-wall AdS/QCD model \cite{EKSS, DP}, and the realization of spin-1/2
nucleons \cite{HIY}. The model has a background geometry as
\begin{eqnarray}
ds^2=G_{MN}\,dx^Mdx^N=\frac{1}{z^2}\,(\eta_{\mu\nu}dx^{\mu}dx^{\nu}-dz^2)\,, \quad  0\leq z\leq z_{\mathrm{IR}}
\end{eqnarray}
This is a slice of $AdS_5$ in the Poincar\'e patch. Here the indices $M=(\mu,5)$ and $\mu=0,1,2,3$.
The convention of Minkowski metric is diag(1,-1,-1,-1). In the limit $m_u=m_d=0$, the two flavor
QCD Lagrangian has $SU(2)_L \times SU(2)_R$ chiral symmetries. The conserved currents are
$\bar{q}_L\gamma^{\mu}t_aq_L$ and $\bar{q}_R\gamma^{\mu}t_aq_R$,
where $q=(u,d)$ and $t_a$ is the generator of $SU(2)$ with $a=1,2,3$. According to the general dictionary
of gauge/gravity duality, there should be two corresponding 5d gauge fields, denoted by $L_{M}^a$ and $R_{M}^a$,
living in the bulk. The operator related to the chiral symmetry breaking is $\bar{q}^{\alpha}_R q^{\beta}_L$,
where $\alpha,\beta$ are indices for the fundamental representation of $SU(2)$. So there should be a scalar
field $X_{\alpha\beta}$ in the bulk which is bifundamental under $SU(2)_L \times SU(2)_R$.
The meson sector can be described by the action
\begin{eqnarray}
S_M=\int d^5x \sqrt{G}\,\, \mathrm{Tr} \left\{|DX|^2-m_X^2|X|^2-\frac{1}{2g_5^2}(F_L^2+F_R^2)\right\}\,.
\end{eqnarray}
Here $F_{L,R}$ is the field strength of $L_M$ and $R_M$, and the covariant derivative of $X$ is
$D_M{X}=\p_M{X}-iL_M{X}+i{X}R_M$. The parameters $m_X^2=-3$ and $g_5^2=12\pi^2/N_c=4\pi^2$ are fixed
by matching with QCD's results. Note that $X$ is tachyonic, and it tends to get a nonzero VEV.
The vector and axial vector gauge fields are defined as $V_M=(L_M+R_M)/\sqrt{2}$ and $A_M=(L_M-R_M)/\sqrt{2}$,
which at the UV boundary $z=0$ couple with the vector and axial vector currents $\bar{q}\gamma^{\mu}t_aq$ and
$\bar{q}\gamma^{\mu}\gamma^5t_aq$ respectively. The classical solution of $X$ is
\begin{eqnarray}
X_0(z)=\frac{1}{2}m_qz+\frac{1}{2}\sigma z^3\,,
\end{eqnarray}
where $m_q$ is the quark mass, and $\sigma$ is the chiral condensate. In the axial gauge $L_5=R_5=0$, the $\rho$
mesons can be identified as Kaluza-Klein modes of $V_\mu(x,z)$, while the $a_1$ mesons are that of $A_\mu(x,z)$.
By setting $X(x,z)=(X_0+S(x,z))\,e^{2i\pi(x,z)}$, the scalar $f_0$ mesons are KK modes of $S(x,z)$. The meson
spectra can be reproduced quite well in this way.

The spin-1/2 nucleons can also been realized in this model. This is done in \cite{HIY} by introducing two 5d
Dirac spinor fields $\Psi_{1,2}$, which transform in the $(\frac{1}{2},0)$ and $(0,\frac{1}{2})$ representation
of $SU(2)_L \times SU(2)_R$. These are the holographical duals of left-handed and right-handed spin-1/2 baryon
operators. The bulk action for the nucleon sector is
\begin{eqnarray}
S_N&=&\int d^5x \sqrt{G}\,\,\mathrm{Tr}(\,\mathcal{L}_K+\mathcal{L}_I) \,, \nonumber\\
\mathcal{L}_K&=&i\bar{\Psi}_1\Gamma^M\nabla_M\Psi_1+i\bar{\Psi}_2\Gamma^M\nabla_M\Psi_2
                -m_5\bar{\Psi}_1\Psi_1+m_5\bar{\Psi}_2\Psi_2 \,, \nonumber\\[0.1cm]
\mathcal{L}\,_I&=&-g\,\bar{\Psi}_1X\Psi_2-g\,\bar{\Psi}_2X^\dag\Psi_1\,.
\end{eqnarray}
Here $\Gamma^M=e^M_A\Gamma^A=z\delta^M_A\Gamma^A$, and $\{\Gamma^A,\Gamma^B\}=2\,\eta^{AB}$
with $A=(a,5)$. We can choose the representation as $\Gamma^A=(\gamma^a, -i\gamma^5)$.
The covariant derivatives for spinors are
\begin{eqnarray}
\nabla_M \Psi_1&=&\p_M\Psi_1+\frac{1}{2}\,\omega^{AB}_M\Sigma_{AB}\Psi_1-iL_M\Psi_1 \,,\\
\nabla_M \Psi_2&=&\p_M\Psi_2+\frac{1}{2}\,\omega^{AB}_M\Sigma_{AB}\Psi_2-iR_M\Psi_2 \,.
\end{eqnarray}
Here $\Sigma_{AB}=\frac{1}{4}[\Gamma_A,\Gamma_B]$, and the nonzero components of the spin
connection $\omega^{AB}_M$ is $\omega^{a5}_\mu=-\omega^{5a}_\mu=\frac{1}{z}\,\delta^a_\mu$.
The mass $m_5$ can be related to the dimension of the baryon operator which may have nonzero
quantum corrections, so we treat $m_5$ as a free parameter.
Note that the 5d spinor does not have chirality, and the 4d chirality is encoded in the sign
of the 5d mass term. For $\Psi_1$ (positive mass) only the right-handed components survive
near the UV boundary $z=0$, which correctly couples with the left-handed baryon operator;
and vice versa for $\Psi_2$. The Yukawa coupling between $\Psi$'s and $X$ is required by
the chiral symmetry breaking in the nucleon sector. It will lift the zero modes and breaks
the spectrum degeneracy of $\Psi_1$ and $\Psi_2$ into a parity doublet pattern. The masses of
first three spin-1/2 nucleons, i.e. $N(939),\,N(1440)$ and $N(1535)$, can be reproduced well
in this model. However when considering higher states, the theoretical values are quite
beyond the experimental data. Actually it is not difficult to see why this happens. Since
the IR cutoff gives a potential just like infinitely deep square-well, the eigenvalues $m_n^2$
grow as $n^2$, while the nucleon masses grow linearly with $n$.

Our observation is that the Yukawa coupling term will give a potential proportional to $z^4$.
If we take a relatively large IR cutoff such that the intersecting point of the $z^4$-potential curve
and the IR wall is higher than 4.41GeV$^2$, i.e. the mass square of $N(2100)$, the low-lying nucleon spectra will be
determined mainly due to the effect of the quartic potentials. Then the growth of the mass spectra is slowed down.
In the next section we will show that all of seven spin-1/2 nucleon
masses listed in PDG can be reproduced quite precisely with errors smaller than 11\%.

\section{Nucleon spectra}

Nucleons in this model are identified as Kaluza-Klein modes of the 5d Dirac spinors
\begin{eqnarray}
\Psi_1(x,z)=\begin{pmatrix} \sum_n N_{1L}^{(n)}(x)f_{1L}^{(n)}(z)\,\, \\[0.2cm] \sum_n N_{1R}^{(n)}(x)f_{1R}^{(n)}(z)\,\,  \end{pmatrix}\,\,;\quad
\Psi_2(x,z)=\begin{pmatrix} \sum_n N_{2L}^{(n)}(x)f_{2L}^{(n)}(z)\,\, \\[0.2cm] \sum_n N_{2R}^{(n)}(x)f_{2R}^{(n)}(z)\,\,  \end{pmatrix}\,\,.
\end{eqnarray}
Here $N^{(n)}$ are two-component Weyl spinors, and $f^{(n)}$ are complex scalar functions.
The internal components $f^{(n)}$ should satisfy \cite{HIY}
\begin{eqnarray}
\begin{pmatrix} \p_z-\frac{\Delta^+}{z} & -v(z) \\[0.2cm] -v(z) & \p_z-\frac{\Delta^-}{z} \end{pmatrix}
\begin{pmatrix} f_{1L}^{(n)} \\[0.2cm] f_{2L}^{(n)} \end{pmatrix}=
-m_{(n)}\begin{pmatrix} f_{1R}^{(n)} \\[0.2cm] f_{2R}^{(n)} \end{pmatrix}\,\,, \label{EOM1}\\[0.3cm]
\begin{pmatrix} \p_z-\frac{\Delta^-}{z} & v(z) \\[0.2cm] v(z) & \p_z-\frac{\Delta^+}{z} \end{pmatrix}
\begin{pmatrix} f_{1R}^{(n)} \\[0.2cm] f_{2R}^{(n)} \end{pmatrix}=
+m_{(n)}\begin{pmatrix} f_{1L}^{(n)} \\[0.2cm] f_{2L}^{(n)} \end{pmatrix}\,\,. \label{EOM2}
\end{eqnarray}
Here $\Delta^{\pm}=2\pm m_5$ and $v(z)=gX_0(z)/z$. To get the correct chiral coupling between $\Psi$'s
and baryon operators at the UV boundary, we need to impose boundary conditions \cite{HIY}
\begin{eqnarray}
f_{1L}^{(n)}(0)=0\,,\quad\quad f_{1R}^{(n)}(z_{\mathrm{IR}})=0\,;\nonumber\\
f_{2R}^{(n)}(0)=0\,,\quad\quad f_{2L}^{(n)}(z_{\mathrm{IR}})=0\,.\label{bcf}
\end{eqnarray}
We can eliminate $(f_{1R}^{(n)},f_{2R}^{(n)})^T$ from EOM (\ref{EOM1}) and (\ref{EOM2}) as
\begin{eqnarray}
&&\begin{pmatrix} f_{1L}^{(n)''} \\[0.2cm] f_{2L}^{(n)''} \end{pmatrix}+
\begin{pmatrix} -\frac{4}{z} & 0 \\[0.2cm] 0 & -\frac{4}{z} \end{pmatrix} \begin{pmatrix} f_{1L}^{(n)'} \\[0.2cm] f_{2L}^{(n)'} \end{pmatrix} \nonumber\\[0.2cm]
&&\quad\quad +\begin{pmatrix} m_{(n)}^2-\frac{m_5^2-m_{_5}-6}{z^2}-v^2 & -v' \\[0.2cm] -v' & m_{(n)}^2-\frac{m_5^2+m_{_5}-6}{z^2}-v^2 \end{pmatrix}
\begin{pmatrix} f_{1L}^{(n)} \\[0.2cm] f_{2L}^{(n)} \end{pmatrix}=0\,.
\end{eqnarray}
We can transform the above equation to a coupled Schr\"{o}dinger form. Define
$(f_{1L}^{(n)},f_{2L}^{(n)})=(z^2\chi_{1L}^{(n)},\,z^2\chi_{2L}^{(n)})$ and
$\chi^{(n)}_L=(\chi_{1L}^{(n)},\chi_{2L}^{(n)})^T$, then
\begin{eqnarray}
\chi_L^{(n)}{''}+(m_{(n)}^2-V(z))\,\chi^{(n)}_L=0\,,\label{Sch}
\end{eqnarray}
where the potential matrix $V(z)$ is
\begin{eqnarray}
V(z)=\begin{pmatrix} V_{11} & V_{12} \\[0.2cm] V_{21} & V_{22} \end{pmatrix}
    =\begin{pmatrix} \frac{m_5^2-m_{_5}}{z^2}+v^2 & v' \\[0.2cm] v' & \frac{m_5^2+m_{_5}}{z^2}+v^2 \end{pmatrix}\,.\label{ptmat}
\end{eqnarray}
The boundary condition for $\chi^{(n)}_L$ can be deduced from that of $f^{(n)}_L$ in (\ref{bcf})
and the EOM (\ref{EOM1}) and (\ref{EOM2}) as
\begin{eqnarray}
\chi^{(n)}_{1L}=0\,,\quad&& \frac{m_5}{z}\chi^{(n)}_{2L}+\chi^{(n)}_{2L}{'}=0\,,\quad   \mathrm{at}\,\, z\rightarrow 0 \,;\label{bc1}\\[0.2cm]
\chi^{(n)}_{2L}=0\,,\quad&& \frac{m_5}{z}\chi^{(n)}_{1L}-\chi^{(n)}_{1L}{'}=0\,,\quad   \mathrm{at}\,\, z\rightarrow z_{\mathrm{IR}}\,.\label{bc2}
\end{eqnarray}
In the chiral symmetric phase, i.e. $v=0$, (\ref{Sch}) becomes two decoupled equations,
and can be solved in terms of Bessel functions. Each eigenvalue has degree 2 of degeneracy.
After turning on the chiral symmetry breaking, the off-diagonal elements will lead to spectrum split
and this is consistent with the parity doublet pattern observed in the nucleon spectra.

The eigenvalues can be numerically calculated. For example in \cite{HIY}, the authors chose
\begin{eqnarray}
m_5=\frac{5}{2}\,,\quad g=14.4\,,\quad z_{\mathrm{IR}}=4.878\,\mathrm{GeV}^{-1}\simeq(0.205\,\mathrm{GeV})^{-1}\,.
\end{eqnarray}
Then the model can reproduce the masses of first three spin-1/2 nucleon states quite well:
$m_{(0)}=0.94\mathrm{GeV},\, m_{(1)}=1.44\mathrm{GeV},\, m_{(2)}=1.50\mathrm{GeV}\,$.
However for higher states the theoretical values are not good enough, e.g. $m_{(3)}=2.08\,\mathrm{GeV}$.
\begin{figure}
\centering
\includegraphics{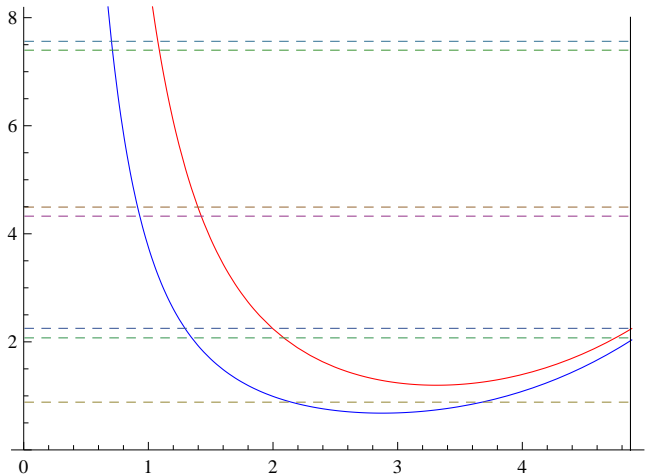}
\caption{\small{The potentials $V_{11},V_{22}$ and eigenvalues (i.e. mass squares)
of \cite{HIY} in the unit of GeV$^2$. The horizontal axis is for the $z$ coordinate.
The effect of the IR wall is notable, roughly an infinite square-well.}}\label{eps.pt1}
\end{figure}
The reason for the overestimation is not difficult to understand.
The intersecting points of $V_{11},V_{22}$ with the IR wall $z=z_{\mathrm{IR}}=4.878\mathrm{GeV}^{-1}$
are low, about 2GeV$^2$ (see Figure \ref{eps.pt1}). So the effect of the IR wall to the eigenvalues is notable,
which acts approximately as a infinitely deep square-well. The eigenvalues $m_{(n)}^2$ grow roughly as $n^2$.
However the experimental data show a Regge behavior, i.e. $m_{(n)}^2\sim n$.

Our observation is that the Yukawa coupling contributes a $v^2$ term in $V_{11}$ and $V_{22}$ potentials
(see (\ref{ptmat})), where
\begin{eqnarray}
v(z)=gX_0(z)/z=\frac{g}{2}(m_q+\sigma z^2)\,.
\end{eqnarray}
Therefore $V_{11}$ and $V_{22}$ both behave like $\frac{1}{4}g^2\sigma^2z^4$ when $z$ is large.
If we choose an IR cutoff such that the intersection between the potential
curves with the IR wall is much larger than $4.41\mathrm{GeV}^2$, i.e. the mass square of $N(2100)$,
the effects of the $z^4$-growth potentials, instead of the hard IR wall, will dominate in the
determination of the low-lying eigenvalues. This will slow down the growth of the mass spectra
and give us a better fit to the experimental data.

In the following discussion we will take the chiral limit, i.e. $m_q=0$. From the analysis of the meson sector,
two parameters $z_{\mathrm{IR}}$ and $\sigma$ are related as \cite{DP}
\begin{eqnarray}
\sigma=\frac{\sqrt{2}\,\,\xi}{g_5z_{\mathrm{IR}}^3}\,,\quad \mathrm{with} \quad \xi=4\,.
\end{eqnarray}
We take $z_{\mathrm{IR}}$ as a free parameter. Although the mass $m_5$ of the 5d Dirac spinors is
related to the dimension of the corresponding baryon operators, we also treat it as a free parameter
since the baryon operator may have nonzero anomalous dimension. In total we have three free parameters
$(m_5,g,z_{\mathrm{IR}})$. We use the shooting method to numerically solve (\ref{Sch}) with boundary
condition (\ref{bc1}) and (\ref{bc2}), and then determine the values of $(m_5,g,z_{\mathrm{IR}})$ by
fitting $m_{(n)}$ with the masses of seven spin-1/2 nucleons.
The trick is firstly to fit the data under the constraint that the intersections between the potential curves
with the IR wall are roughly 4 GeV$^2$, and then using the result as the initial guess to do a real three-parameter
fitting. The result is
\begin{eqnarray}
m_5=3\,,\quad g=32.8\,,\quad z_{\mathrm{IR}}=7.663\mathrm{GeV}^{-1}\simeq(0.1305\mathrm{GeV})^{-1}\,. \label{para}
\end{eqnarray}
The resulting mass spectra are listed in Table \ref{mass}.
\begin{table}
\centering
\begin{tabular}{|c|c|c|c|c|c|c|c|}
\hline
$n$                &  0   &   1  &  2   &  3   &  4   &  5   &   6  \\
\hline
$m_{\mathrm{exp}}$ & 0.94 & 1.44 & 1.54 & 1.65 & 1.71 & 2.09 & 2.10 \\
\hline
$m_{\mathrm{th}}$  & 0.98 & 1.29 & 1.40 & 1.66 & 1.70 & 2.01 & 2.02 \\
\hline
error              &4.3\% &10.4\%&9.1\% &0.6\% &0.6\% &3.8\% &3.8\% \\
\hline
\end{tabular}
\caption{\small{The experimental and theoretical values of the spin-1/2 nucleon masses.
The parameters are $m_5=3,\, g=32.8,\, z_{\mathrm{IR}}=7.663\mathrm{GeV}^{-1}
\simeq(0.1305\mathrm{GeV})^{-1}$, and $\sigma=\frac{4\sqrt{2}}{g_{_5}z_{\mathrm{IR}}^3}
=(0.126\mathrm{GeV})^3$. The average error is remarkably 4.66\%.}}\label{mass}
\end{table}
\begin{figure}
\centering
\includegraphics{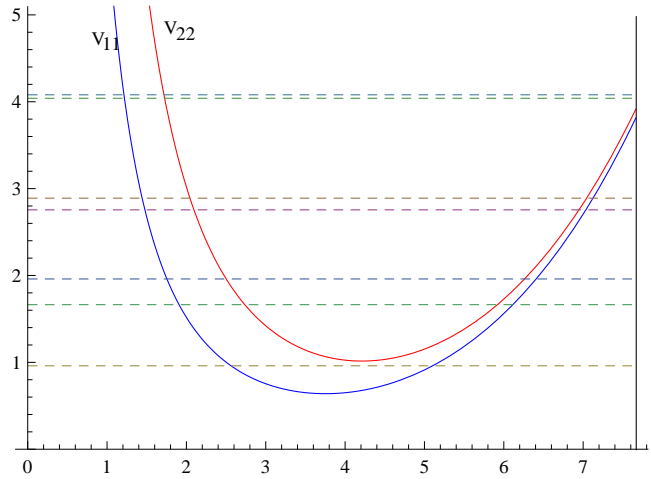}
\caption{\small{The potentials $V_{11},V_{22}$ and eigenvalues (i.e. mass squares) in the unit of GeV$^2$.
The horizontal axis is for the $z$ coordinate.
The parameters are $m_5=3,\, g=32.8,\, z_{\mathrm{IR}}=7.663\mathrm{GeV}^{-1}$
and $\sigma=(0.126\mathrm{GeV})^3$.}}\label{eps.pt2}
\end{figure}
\hspace{-0.3cm}It is remarkable that the errors are all smaller than 11\% for seven nucleon states.
The average error is 4.66\%.
In Figure \ref{eps.pt2} we plot the potentials $V_{11}$ and $V_{22}$, together with seven eigenvalues,
i.e. mass squares. We see that, for low-lying eigenvalues, the effects of the $z^4$-growth
potentials are dominant, while the IR wall can hardly be felt by these nucleon states. Readers can
see the difference clearly by comparing Figure \ref{eps.pt1} with Figure \ref{eps.pt2}.

\section{Pion-Nucleon coupling}

In this section we use the parameters determined in the last section to calculate the pion-nucleon coupling.
We will follow the method presented in \cite{HIY}. In the chiral limit, the Kaluza-Klein wave function $f_0$
for the pion field can be analytically obtained \cite{KKY}
\begin{eqnarray}
f_0(z)=N_0 z^3\left(\frac{I_{2/3}(g_5\sigma z^3_{\mathrm{IR}}/3)}{\,I_{-2/3}(g_5\sigma z^3_{\mathrm{IR}}/3)}\,I_{-2/3}(g_5\sigma z^3/3)
      -I_{2/3}(g_5\sigma z^3/3)\right)\,,
\end{eqnarray}
where $I_{\pm2/3}$ are the 1st kind modified Bessel function. This is the unique solution of the boundary problem
\begin{eqnarray}
\p_z\left[\frac{z^3}{X_0^2}\,\,\p_z\left(\frac{f_0}{z}\right)\right]-4g_5^2f_0=0\,,\qquad
f_0(0)=f_0(z_{\mathrm{IR}})=0\,.
\end{eqnarray}
The constant $N_0$ is fixed by the normalization condition
\begin{eqnarray}
\int_0^{z_{_{\mathrm{IR}}}}dz \left[\,\frac{f_0^2}{2g_5^2z}+\frac{z^3}{8g_5^4X_0^2}\left(\p_z\left(\frac{f_0}{z}\right)\right)^2\right]=1\,.
\end{eqnarray}

The coupling between the pion and nucleons comes from the covariant derivative terms of the bulk fermions
and the Yukawa coupling terms. The 4d coupling constant $g^{(n)}_{\pi NN}$ can be written \cite{HIY} as an integration along the $z$-direction as
\begin{eqnarray}
g^{(n)}_{\pi NN}=\frac{1}{2\sqrt{2}}\,\int_0^{z_{_{\mathrm{IR}}}}\frac{dz}{z^4}
   \left[f_0\left(\bar{f}_{1L}^{(n)}f_{1R}^{(n)}-\bar{f}_{2L}^{(n)}f_{2R}^{(n)}\right)-
   \frac{gz^2}{\,2g_5^2X_0}\left(\bar{f}_{1L}^{(n)}f_{2R}^{(n)}-\bar{f}_{2L}^{(n)}f_{1R}^{(n)}\right)\right],
\end{eqnarray}
with the normalization condition
\begin{eqnarray}
\int_0^{z_{_{\mathrm{IR}}}} \frac{dz}{z^4}\left(|f_{1L}^{(n)}|^2+|f_{2L}^{(n)}|^2\right)=
\int_0^{z_{_{\mathrm{IR}}}} \frac{dz}{z^4}\left(|f_{1R}^{(n)}|^2+|f_{2R}^{(n)}|^2\right)=1\,.
\end{eqnarray}
Note that for parity-even states we have $f_{1L}^{(n)}=f_{2R}^{(n)}$ and $f_{2L}^{(n)}=-f_{1R}^{(n)}$,
while for parity-odd states we have $f_{1L}^{(n)}=-f_{2R}^{(n)}$ and $f_{2L}^{(n)}=f_{1R}^{(n)}$.
The coupling between pion and the $n=0$ state, i.e. the (p, n) doublet, is measured as $g^{(0)}_{\pi NN}\simeq13$.
By using the parameters listed in (\ref{para}), the calculated value is
\begin{eqnarray}
g^{(0)}_{\pi NN}\simeq11\,.
\end{eqnarray}

\section{Summary}

In this paper we reconsider the spin-1/2 nucleon mass spectra in a hard-wall AdS/QCD model \cite{HIY}.
By introducing two Dirac spinors in the bulk, the nucleons can be identified as their Kaluza-Klein
modes along the extra fifth dimension. Chiral symmetry breaking is included through a Yukawa
coupling term, which gives a potential growing as $z^4$ when $z$ large. Our observation is that,
if we choose a relatively large IR cutoff such that the intersection between the potential curves
and the IR wall is high enough, the low-lying states will mainly feel the effect of the $z^4$-growth
potential, while the IR wall is almost screened. This will slow down the growth of the mass spectra,
and give us a better fitting to the experimental data. After taking the chiral limit $m_q=0$, we choose our
three parameters as $m_5=3,\, g=32.8,\, z_{\mathrm{IR}}=7.663\mathrm{GeV}^{-1}\simeq(0.1305\mathrm{GeV})^{-1}$.
The resulting mass spectra fit the data quite well, see Table \ref{mass}. Remarkably, all errors are smaller than 11\%
for seven spin-1/2 nucleons listed in PDG, and the average is only 4.66\%. We also calculate the pion-nucleon
coupling constant. The result is $g_{\pi NN}\simeq11$, which also improves early results in the literature.

By the results of this paper, it seems that extending this nucleon
model to the soft-wall AdS/QCD is a very natural next step
\footnote{Actually there are already some works in this direction,
see e.g. \cite{BdT(N)}.}. The quadratic growth dilaton field plays
the role of a soft-wall in the meson sector. However in the nucleon
sector the dilaton cannot do that job, since the action of Dirac
spinors is first order. A proper choice of the bulk scalar VEV
may offer a needed soft-wall through the Yukawa coupling. This is
now under investigation \cite{Z}.

\section*{Acknowledgements}

I would like to thank Prof. Y.-C. Huang and W.-Y. Wang for many interesting discussions,
and Prof. Y.-L. Wu for his talk given in the workshop on particle physics at BJUT, which
stimulated the present work.

\end{document}